\documentclass[12pt,preprint]{aastex}
  \slugcomment{Submitted to ApJ, 2005 Oct. 21; Revised 2005 Dec. 23 and 2006 Jan. 20}
\shorttitle{Properties of Binary KBO (47171) 1999 TC$_{36}$}
\shortauthors{Stansberry et al.}

\newcommand{\Spitzer}{{\it Spitzer}}
\newcommand{\tc}{(47171) 1999~TC$_{36}$}
\newcommand{\eg}{{\it e.g.}}
\newcommand{\etal}{{\it et al}.}
\newcommand{\mum}{$\mu$m}
\newcommand{\gcc}{g/cm$^3$}

\begin{document}


\title{The Albedo, Size, and Density of Binary Kuiper Belt Object (47171) 1999 TC$_{36}$}


\author{J.A. Stansberry\altaffilmark{1}}
\email{stansber@as.arizona.edu}
\author{W.M. Grundy\altaffilmark{2}}
\author{J.L. Margot\altaffilmark{3}}
\author{D.P. Cruikshank\altaffilmark{4} }
\author{J. P. Emery\altaffilmark{5}}
\author{G.H. Rieke\altaffilmark{1}}
\author{D.E. Trilling\altaffilmark{1}}

\altaffiltext{1}{University of Arizona, Tucson, AZ 85721}
\altaffiltext{2}{Lowell Observatory, Flagstaff, AZ 86001}
\altaffiltext{3}{Cornell University, Ithaca, NY 14853}
\altaffiltext{4}{NASA Ames Research Center, Moffett Field, CA 94035}
\altaffiltext{5}{SETI Institute, Mountain View, CA 94043}
%\altaffiltext{6}{Institute for Astronomy, University of Hawaii, Honolulu, HI 96822}
%\altaffiltext{7}{Jet Propulsion Laboratory, Pasadena, CA 91109}

\begin{abstract} We measured the system-integrated thermal emission of
the binary Kuiper Belt Object \tc\ at wavelengths near 24 and 70 \mum\
using the \Spitzer\ space telescope. We fit these data and the visual
magnitude using both the Standard Thermal Model and thermophysical
models. We find that the effective diameter of the binary is 405~km, with
a range of 350 -- 470~km, and the effective visible geometric albedo for
the system is 0.079 with a range of  0.055 -- 0.11.  The binary orbit,
magnitude contrast between the components, and system mass have been
determined from HST data (Margot et al., 2004; 2005a; 2005b).  Our effective
diameter, combined with that system mass, indicate an average density
for the objects of 0.5~\gcc, with a range 0.3 -- 0.8~\gcc.  This density
is low compared to that of materials expected to be abundant in solid
bodies in the trans-Neptunian region, requiring 50 -- 75\% of the interior
of \tc\ be taken up by void space.  This conclusion is not greatly affected 
if \tc\ is ``differentiated'' (in the sense of having either a rocky or just
a non-porous core). If the primary is itself a binary, the average density 
of that (hypothetical) triple system would be in the range 0.4 -- 1.1~\gcc, 
with a porosity in the range 15 -- 70\%. 
\end{abstract}


\section{Introduction}

The physical properties of Kuiper Belt Objects (KBOs) remain poorly known
13 years after the discovery of the first KBO, (15760) 1992~QB$_1$ (Jewitt
and Luu, 1993). While they can be discovered, their orbits determined,
and their visible-light colors measured (to some extent) using modest
telescopes, learning more about KBOs generally requires the largest of
ground-based telesopes, or space-based instrumentation. The {\it Hubble
Space Telescope} (HST) has revealed that the Kuiper Belt hosts a large
population of binary systems (\eg\ Noll, 2003).  Currently 21 KBOs are
known to be binary (Veillet \etal\  2002; Elliot \etal\ 2001; Noll \etal\
2002a, 2002b, 2002c; Millis \& Clancy 2003; Stephens \& Noll, 2004;
Stephens and Noll, 2005; Cruikshank \etal\ 2005). Very recently, two
new moons were discovered in the Pluto-Charon system (Weaver \etal\ 2005;
Buie \etal\ 2005), making it a quadruple, and the large KBO 2003~EL$_{61}$
has been shown to have 2 moons, making it a triple (Brown \etal\ 2005).

Stephens and Noll (2005) find that of the 81 KBOs they observed with
HST/NICMOS, 11\% are binary (at current detection limits). They also
suggest find that objects with very low inclinations and eccentricities
(``cold-classical'' objects) are likely to be binary at a rate of 21\%,
while other, dynamically hotter classes, are binary at a rate of 6\%. The
high incidence of binarity among KBOs is of considerable intrinsic
interest as a probe of the dynamics of the Kuiper Belt (\eg\ Stern,
2002; Goldreich \etal\ 2002; Astakhov \etal\ 2005).
%% Following was removed at the dumbass insistence of the referee.
%  In contrast, there
%is an apparent dearth of binaries among the Centaur population (about
%6 binary Centaurs could already have been discovered if 11\% of Centaurs
%were binary).  The lack of known binary Centaurs may be entirely due to 
%the lack of dedicated, high angular resolution searches such as have 
%been conducted for KBOs.  If there is indeed a dearth
%of binary Centaurs, the discrepancy may provide a clue to the mechanism
%by which KBOs become Centaurs, given that the Centaurs are thought to be
%the dynamical progeny of KBOs.  
However, binaries are also of particular
interest because their masses can be determined from observations of the
relative motions of the components. Such measurements would otherwise be
impossible without sending a spacecraft. When combined with measurements
of sizes, the masses of primitive solar system objects can be used to
constrain their density. The densities of KBOs are indicative of their
interior structure, and can provide clues to the composition of and
conditions in the outermost primitive solar nebula.

\Spitzer\ (Werner \etal\ 2004) offers unique capabilities for measuring
the infrared emission from KBOs and other solar system objects, thereby
constraining their albedos and sizes (\eg\ Stansberry \etal\ 2004;
Cruikshank \etal\ 2004; Lisse \etal\ 2005).  \tc\ and three other KBOs
have been detected in the sub-millimeter (Altenhoff \etal\ 2004; Margot
\etal\ 2002; Lellouch \etal\ 2002; Jewitt \etal\  2001; see also Grundy
\etal\ 2005). One KBO, Quaoar, has been resolved at optical wavlengths,
providing an estimate of its size (Brown and Trujillo 2004).  For KBOs,
the Multiband Imaging Photometer for \Spitzer\ (MIPS), with photometry
channels near 24, 70, and 160 \mum, covers the peak in the typical
spectral energy distribution near 80 \mum.  Measurements at these
wavelengths are sensitive to the distribution of temperature across
the surface of a KBO, and so can reveal something about the thermal
properties of near-surface layers as well as constraining the size and
albedo of the target.

\subsection{(47171) 1999 TC$_{36}$}

Here we report on our \Spitzer\ observations of the binary KBO \tc, the 
albedo and size that are indicated by those data, and the density
we derive by combining our results with the {\it Hubble Space Telescope}
(HST) visible-imaging results of Margot \etal\  (2004; 2005a; 2005b).
\tc\ orbits in the 3:2 resonance with Neptune, and so likely formed
closer to the Sun than its current semi-major axis would suggest (\eg\
Malhotra, 1993). It is very red, with $B-V \simeq 1.05$, $V-R \simeq 0.70$
(Dotto \etal\  2003; Doressoundiram \etal\  2005) and $V-J \simeq 2.3$
(Dotto \etal\  2003; McBride \etal\  2003). Lazzarin \etal\  (2003)
and Dotto \etal\  (2003) give visible to near-IR spectra, and Dotto
suggests a combination of tholins, carbon and water ice as a plausible
surface composition. There appears to be some slight confusion as to
the absolute magnitude, $H_V$, of this object. Both the IAU Minor
Planet Center (MPC, http://cfa-www.harvard.edu/iau/mpc.html) and
the Solar System Dynamics Division at the Jet Propulsion Laboratory
(Horizons, http://ssd.jpl.nasa.gov/horizons.html) give values of
$H_V < 5$, presumably based on apparent magnitudes from astrometric
observations. Photometric studies of \tc\ all conclude that $H_V$ is in
the range 5.33 -- 5.55 (Margot \etal\  2005a; 2005b; Tegler and Romanishin 2005;
Doressoundiram \etal\ 2005; McBride \etal\ 2003, Delsanti \etal\  2001; Boehnhardt
\etal\ 2001).  Here we adopt the value of Doressoundiram \etal\
(2005), $H_V = 5.37 \pm 0.05$, as representative of the range of $H_V$.
The $\leq 5$\% scatter in the $H_V$ determinations by different groups is
small relative to the photometric errors of our \Spitzer\ measurements,
so we do not track its quite small contribution to the uncertainties
in our results.  (We also note a systematic bias between the MPC and
Horizons absolute magnitudes, and absolute magnitudes found in dedicated
photometric surveys, as detailed by Tegler and Romanishin (2005).)

Trujillo and Brown (2002) identified \tc\ as a binary, with a separation
of about 8000 km. Margot \etal\ (2005a; 2005b) imaged the binary with
HST over a period of 25 months, determining the system's orbital period,
semi-major axis, and total mass, to high precision. Table~1 summarizes
the heliocentric orbit of \tc, and the binary parameters of the system as
determined in that study.  In their PSF and orbital fits for \tc, Margot
et al found that the residuals were significantly higher than for other
binaries in their sample, a finding that can possibly be attributed to
the primary being itself a binary, an idea we return to later
%XXX Jean-Luc requested a change of this section
%Interestingly, they also found that the visible colors of the primary
%and secondary were very similar, a feature of all 5 KBO binary systems
%they imaged with HST.  In analyzing the astrometry for \tc\ Margot \etal\
%(2005a; 2005b) found that the solutions for its orbit were significantly
%more uncertain than those for the other 4 objects they had data for,
%and that the larger uncertainties were probably due to assymetries in
%the point spread function of light from the primary.  This observation
%is consistent with the primary itself being a binary, an idea we return
%to later.XXX

\section{Observations and Data Analysis}

We observed \tc\ with the Multiband Imaging Photometer for \Spitzer\
(MIPS, Rieke \etal\ 2004) in its 24 and 70 \mum\ bands, which have
effective wavelengths of 23.68 and 71.42~\mum. Data were collected using
the photometry observing template, which is tailored for photometry of
point sources.  The telescope tracked the target during the observations,
although the motion was negligible relative to the size of the \Spitzer\
point-spread function (PSF) at these wavelengths. In photometry mode
images of the target are taken at many dithered positions on the
arrays to improve the photometric accuracy and the sampling of the
PSF. Photometric repeatability on moderately bright sources is better
than 2\% at 24 \mum, and is 5\% at 70 \mum. The uncertainty in the
absolute calibration of these bands is 5\% and 15\% respectively. For
purposes of fitting models to our photometry, we use uncertainties that
are the root-square-sum of the absolute calibration uncertainties and
the measurement uncertainties determined from the images themselves.
The widths of the filter bandpasses are about 25\%, resulting in modest
color corrections. We iteratively applied color corrections to our
photometry, which converged to give a color temperature of 64.5~K, and
color corrections of $+1$\% and $+10$\% at 24 and 70~\mum\ respectively.
The uncertainties on the correction factors are perhaps a few percent
of the factors themselves, and so are negligible for our study.
Color corrections for the MIPS bands are available from the {\it Spitzer
Science Center} website (http://ssc.spitzer.caltech.edu).

We reduced the raw data and mosaicked them using the MIPS instrument team
data analysis tools (Gordon \etal\ 2004).  For the 24~\mum\ data basic
processing included slope fitting, flat-fielding, and corrections for
droop. All of these steps are currently implemented in the {\it Spitzer
Science Center} (SSC) pipeline products.  Additional corrections were made
to remove readout offset (a jailbar pattern in the images), the effects
of scattered light (which introduces a pointing-dependent background
gradient and slightly degrades the sensitivity), and the application of
a second-order flat, derived from the data itself, to remove latents
from previous observations. These additional processing steps will be
incorporated into the SSC pipeline data products by {\it circa} the end
of 2006. For the field where we observed \tc, the additional processing
reduced the total noise (including both conventional noise and that due
to background sources) in the 24~\mum\ image by 30\%.  The 70~$\mu$m
data processing was basically identical to that of the {\it Spitzer}
pipeline (version S12).  We converted the calibrated instrumental count
rates to flux density using factors of 1.042 $\mu$Jy/$\arcsec^2$ and
16.6 mJy/$\arcsec^2$ in the 24 and 70 $\mu$m bands, respectively. 
Figure~1 shows the resulting images.

We measured the flux density of \tc\ using apertures $9\farcs96$
and $29\farcs6$ (about 4 and 3 native pixels) in diameter at 24 and
70~\mum; the PSF full-width at half-max is $6\farcs5$ and $20\arcsec$
in those bands.  The apertures were positioned at the center-of-light
centroid.  These small apertures improve the signal-to-noise ratio
of the measurements, but require both excellent sampling of the PSF,
and the application of significant aperture corrections.  Mosaics were
constructed using $1.245\arcsec$ and $4.925\arcsec$ pixels at 24 and
70~$\mu$m (about $1/2$ the native pixel scale of those arrays). Mosaicking
at this level of sub-sampling, in combination with the multiply-dithered
nature of the observations, results in images with excellent PSF sampling
and centroid reproducibility.  We computed aperture corrections using
smoothed STinyTim model PSFs. We smoothed the model PSFs until they agreed
with observed stellar PSFs, giving a noiseless, perfectly normalized,
zero-background model PSF.  We performed photometry on the result,
deriving aperture corrections of 1.91 and 1.85 at 24 and 70~\mum,
respectively. This method has been verified to result in errors $\leq
1$\% in the photometry of calibration stars of moderate brightness.
The response of the MIPS detectors is quite linear for signals ranging
from the natural sky background to near the saturation limit, so our
aperture corrections should apply accurately for photometry of faint
sources.  Table~2 summarizes the circumstances of our observations and
the measured, color-corrected flux densities: $1.23\pm 0.06$ and $25.4\pm
4.7$~mJy at 24 and 70~\mum, respectively.

Because we did not make a second observation of the field where we
detected \tc, it is impossible to rule out the possibility that a
background source may have coincided with our target, biasing our
photometry. However, the areal density of extra-galactic sources bright
enough to have influenced our results is quite low, and the probability
that one could have serendipitously fallen within our beam is quite low
in both bands. The areal density of 0.1$\pm$0.05 mJy 24 $\mu$m sources
is 3 / arcmin$^2$ (Papovich \etal\ 2004). For our aperture size (0.022
arcmin$^2$) this translates into a probability of 6.6\% that our 24~\mum\
photometry could be contaminated by a 0.1 mJy source. The probability that
a fainter source falls in the beam is higher, but such a source would have
an insignificant effect on our photometry.  We compute the areal density
of 70 $\mu$m sources based on the results of Dole \etal\ (2004). At
70 $\mu$m the density of $3\pm1$ mJy sources is $\sim0.4$/arcmin$^2$,
which translates to an 7.6\% probability of a contaminating source 
in our 0.19 arcmin$^2$ aperture. Again, fainter sources are more likely than
this to fall within the beam, but would have an insignificant effect on
the photometry.  Thus, not only is the probability of contamination by
an extra-galactic source small at relevant flux densities, the effect on
our photometry would be small relative to our estimated flux errors.
Furthermore, the observed 24 to 70um color is redder than would be
expected for a typical background source. For \tc\, the 70:24 ratio
of fluxes is $\sim20$, whereas the results of Papovich et al. (2004)
and Dole et al. (2004), indicate that a typical ratio for a background
galaxy is $\sim10$. Thus is is unlikely that our photometry is
strongly contaminated by a background galaxy, because such contamination
would result in an untenably high color temperature, whereas the color
temperature we derive (see below) is very plausible for an object at
the distance of \tc.

Sources within the galaxy can also contaminate far-IR photometry. In
particular, thermal emission from extended, uneven clouds of interstellar
dust can increase the noise in regions of the sky where such ``infrared
cirrus'' appears in the background. An advantage of \Spitzer\ and MIPS
over earlier far-IR missions is the relatively small PSF, the fact that
the PSF is well-sampled by the pixels, and the fact that the field of view
of the arrays is significantly larger than the PSF.  These facts allow
direct estimation of the contribution of IR cirrus (and extra-galactic
confusion for that matter) directly from the images. Inspection of
Figure~1 shows that cirrus did not constribute significantly to our
photometric errors.

It also seems unlikely that there could be a coma contributing
significantly at 24 or 70~\mum, because there is no evidence for
variability in the visible colors or absolute magnitudes from multiple
photometric observations.  A possibility that we can not rule out is
that \tc\ posesses a coma dominated by relatively large particles that
would scatter poorly at visible to near-IR wavelengths, but which could
contribute significantly in the thermal. Given the lack of any direct
evidence of a coma around \tc, and the requirement that it be both
IR-bright and V-invisible to affect our interpretation of the data,
we do not pursue this possibility further here.

\subsection{Thermal Modeling and Results}

The Standard Thermal Model (STM, Lebofsky \& Spencer 1989) is the most
widely used model for interpreting observations of thermal emission
from small bodies in the asteroid main belt and the outer Solar System
({\it c.f.} Tedesco et al. 2002; Fern\'{a}ndez et al. 2002; Campins
et al.  1994). The model assumes a spherical body whose surface is in
instantaneous equilibrium with the insolation, equivalent to assuming
either a thermal inertia of zero, a non-rotating body, or a rotating
body illuminated and viewed pole-on.  In the STM the subsolar point
temperature is
\begin{equation}
T_0 = [S_0(1-p_Vq)/(\eta\epsilon\sigma)]^{1/4}\;,
\end{equation}
where $S_0$ is the solar constant at the distance of the body, $p_V$ is
the geometric albedo, $q$ is the phase integral (assumed here to be
0.39, equivalent to a scattering assymetry parameter, $G = 0.15$ (Lumme
and Bowell 1981; Bowell \etal\ 1989)), $\eta$ is the beaming parameter,
$\epsilon$ is the emissivity (which we set to 0.9), and $\sigma$ is the
Stefan-Boltzmann constant. Given $T_0$, the temperature as a function of
position on the surface is $T = T_0 \mu^{1/4},$ where $\mu$ is the cosine
of the insolation angle.  The nightside temperature is taken to be zero.
Surface roughness leads to localized variations in surface temperature
and non-isotropic thermal emission (beaming) because individual points on
the surface radiate their heat preferentially in the sunward direction.
Thus, when viewed at small phase angles, rough surfaces appear warmer than
smooth ones, and the thermal emission tends to be dominated by emission
from the warmer depressions and sunward-facing slopes. This effect is
captured by the beaming parameter, $\eta$. Lebofsky et al. (1986) found
$\eta = 0.76$ for Ceres and Vesta; the nominal range for $\eta$ is 0 to
1, with unity corresponding to a perfectly smooth surface (Lebofsky \&
Spencer 1989).  In modeling the thermal emission from a large sample of
Jovian Trojan asteroids, Fern\'{a}ndez \etal\ (2003) found a typical
value of $\eta$ for that population of about 0.94. It is worth noting
that $\eta$ enters the equations for the surface temperature in a manner
analogous to the emissivity, so our results could be couched in terms
of emissivity rather than beaming parameter.

The thermal emission from a target is calculated by computing a surface
integral of the Planck function over the visible portion of the object
(because we do this integral, our model is formally a modified version
of the STM).  The Planck function at a particular point on the surface
depends on the calculated temperatures (Eq. 1) and the wavelength
of interest.  This flux density is then scaled by a dilution factor
proportional to $D^2/\Delta^2$, where $D$ is the diameter, and $\Delta$
is the \Spitzer-centric distance.  The phase angle, which is invariably
very small for KBOs, enters into the integral over the visible
hemisphere. Here we have set $\alpha=0$, as the effects are negligible for
the actual phase of our observations ($\alpha = 1.86\degr$).  The total
flux density thus depends on both the target's unknown diameter, $D$,
and albedo, $p_V$, as well as its distance from the Sun and the observer.
Solutions for the size and albedo require a second equation; the object's
visible magnitude typically provides this constraint. We used the
absolute visual magnitude of Doressoundiram \etal\  (2005), $H_V = 5.37$,
to relate the diameter and albedo via $D = 10^{-H_V/5} 1329/\sqrt{p_V}$
({\it e.g.,} Harris 1998), where $D$ is the diameter in km.

The 24 and 70~$\mu$m photometry and results from the STM thermal model
are shown in Figure 2a.  Using the values of q and $\epsilon$ noted
above, we allowed $\eta$ to vary.  The best fit to the data in Figure
2a has $D=420$~km, $p_V = 0.073$, and $\eta = 1.2$. The range of model
parameters that are consistent with the $1\sigma$ error bars of our data
are summarized in Table~3. We note that the model parameters are the
{\it system-average, or effective, values of the diameter, albedo, and
beaming parameter}, because the system is not resolved at 24 nor 70~\mum\
(nor at V for the ground-based visible measurements).  It is not unusual
to allow $\eta$ to range above unity (Harris 1998; Fern\'{a}ndez
et al. 2003; Delbo et al. 2003) when fitting thermal data with the STM.
As can be seen from Equation 1, $\eta>1$ will result in overall lower
surface temperatures, even though $\eta$ was traditionally introduced to
model elevated localized temperatures caused by roughness. We interpret
$\eta>1$ to be an indication that the body has a non-zero thermal
inertia, a relatively short rotation period, a fairly smooth surface,
or some combination of the three.  Therefore, the idealized assumptions in
the STM do not perfectly apply. This is not entirely surprising given
the low temperature of the object, which has the effect of lengthening
the radiative cooling timescale, violating the instantaneous thermal
equilibrium assumption of the STM.

The Isothermal Latitude Model (ILM) is the opposing end-member model to
the STM: the target is assumed to have a surface with infinite thermal
inertia, or, equivalently, to rotate instantaneously (also called the
``fast-rotator model"; Lebofsky \& Spencer 1989).  In real terms, the
ILM applies for objects with rotation periods much shorter than the
timescale for radiative cooling of the surface.  It is also typically
assumed that the subsolar latitude is zero, although other geometries can
be readily computed. In this geometry the surface temperature depends
only on the latitude, and is constant over longitude. Strictly speaking, 
under the ILM $\eta=1$. 

Figure 2b compares our photometry with ILM models.  We were unable to
fit the data under the assumption that $\eta=1$, and so again allowed
it to be a free parameter. The resulting best fit has $D=401$ km, 
$p_V = 0.080$, and $\eta=0.44$.  The range of model parameters consistent
with our data are summarized in Table~3. We interpret $\eta<1$ under the ILM as
an indication that the thermal inertia could actually be rather low
(although the STM results indicate that it is very likely greater than
zero).

If the orientation of the rotational axis and the rotational period are
known, it is possible to improve the ILM model results by accounting for
the actual viewing geometry. Such a tilted ILM is an end-member case of
a thermophysical model, incorporating the time-dependence of insolation,
conduction, and re-radiation.  If we assume that the rotational axes
of the \tc\ binary components are approximately aligned with the
orbit normal, the viewing geometry of our observations is specified.
Using the position of the normal to the \tc\ orbit (Margot \etal\
2005b), we find that the sub-\Spitzer\ and sub-Solar latitude was $\simeq
49\degr$. Table~3 summarizes the results of the ILM for this geometry.

Using the same assumption for the orientation of the rotation axes, we
also fit the data using a smooth-surface thermophysical model, which
includes the time-dependent effect of conduction into and out of the
subsurface (\eg\ Spencer 1990).  The results in terms of $p_V$, $D$,
and thermal inertial, $\Gamma$, and for assumed rotational periods of 10
and 40 hours are summarized in Table~3. The spectral energy distributions
resulting from these models are nearly indistinguishable from those shown
in Figure~2.  Also, because we have added a model parameter, somewhat 
larger ranges of $p_V$ and $D$ are found to be consistent with
the data.

We also undertook a Monte Carlo approach to see if we could constrain
the pole orientation of \tc. In this case we used the rough-surface
thermophysical model of Spencer (1990). Thermal inertias (5 --
100), emissivities (0.9 -- 1.0), albedos (0.01 -- 0.5), rotational
periods (6 -- 32 hr), and spin orientations were all varied randomly
(within the ranges just given); the spin orientation was distributed
homogeneously in terms of solid-angle with respect to the line of sight.
Parameters from thermophysical models with 24 and 70 \mum\ fluxes consistent
with our data and the absolute V magnitude for \tc\ (to within $\pm 1\sigma$) 
were tabulated.  Of the 300000 models we ran, 539 were consistent with the 
observations.  The results of this modeling are shown in Figure~3. While
a significant range of model parameters are consistent with our thermal
measurements, clustering of the points shows that some portions of the
parameter space are more likely than others.  Acceptable models
had geometric albedos between 0.052 and 0.108, and effective diameters
between 340 and 483 km, entirely consistent with the ranges of those
parameters found earlier, and confirming that those ranges encompass
uncertainties resulting from model assumptions.

Interestingly, thermophysical models with very high ($>70$\degr) or with
very low ($<24$\degr) sub-solar latitudes could not fit the data. 80\% of
the fitting models have sub-solar latitudes between 40\degr\ and 60\degr,
neatly bracketing the 49\degr\ implied by alignment between the spin-axis
and the satellite orbit angular momentum vector.  Tidal evolution
frequently drives satellite spins towards zero obliquity (Peale, 1999),
but the obliquity of the primary is relatively insensitive to tidal
evolution under the influence of the satellite. If the \tc\ binary
formed in a massive impact (similar to the Pluto-Charon forming impact,
Canup 2005) the angular momenta vectors of the orbit and primary rotation should
be nearly parallel.  Other mechanisms proposed for forming Kuiper Belt
binaries do not appear to predict any particular relationship between the
orientation of spin and orbital angular momenta (Weidenschilling 2002;
Goldreich \etal\ 2002; Funato \etal\ 2005). If our prediction that the
spin of the \tc\ primary is roughly aligned with the orbital angular
momentum could be independently verified (or disproved), it would help
constrain how the binary formed in the first place.

The rough-surface thermophysical models suggest that $\Gamma >
30$~J\,m$^{-2}$\,K$^{-1}$\,s$^{-1/2}$ for \tc. Many of the best fitting
models are clustered at the upper end of the range of $\Gamma$ we explored
(see Figure~3), so it is possible the thermal inertia is greater than
the upper limit of 100 which we imposed.  Some models that are consistent
with the data also have $\Gamma$ as low as 7.5, but are associated with
improbable combinations of the other model paramters. $\Gamma$ is around
50 for the Moon, and in the range 15 -- 100 in these units for Pluto
(Lellouch \etal\ 2000). Cruikshank \etal\ (2004) estimated $\Gamma$
for 2002~AW$_{197}$ to be below 20, for an assumed equatorial viewing
geometry. As the geometry moves to higher latitudes the effect of thermal
inertia is diluted (in the extreme case, a pole-on ILM is quivalent to
the STM, and no constraint can be placed on $\Gamma$). These values of
$\Gamma$ are consistent with a porous or particulate surface, or the
presence of a low-conductivity material such as amorphous water ice or
unconsolidated carbonaceous particles in the surface layers.

From all of these model fits to the data, we find that the effective
diameter of the \tc\ system is in the range $350 < D < 470$ km, and the
visible geometric albedo in the range $0.055 < p_V < 0.11$.  These adopted
values (Table~3) encompass both measurement uncertainties on the fluxes,
and the uncertainties associated with interpreting the fluxes using
the suite of thermal models described above. Because the only common
outputs of all the models are albedo and size, we only provide adopted
values for those quantities.  These parameter ranges exclude some of the
most extreme models that fit the data, but span most of the reasonable
models.  It further appears that the surface has a thermal inertia large
enough to result in non-zero night-side temperatures, but that diurnal
temperature variations are also likely to be important. For STM models,
this is evidenced by the necessity to set $\eta$ to values greater than
unity, while for ILM models we have to set it to values around 0.5 (see
Table~3). The thermophysical models confirm that \tc\ has a temperature
distribution intermediate between the STM and ILM, with thermal inertia
between about 30 and 100~J\,m$^{-2}$\,K$^{-1}$\,s$^{-1/2}$.

\subsection{Comparison with sub-millimeter results.} 

Altenhoff \etal\ (2004) used the 30~m IRAM mm telescope to measure
the 1.2 mm flux density from \tc\ as $1.1 \pm 0.26$~mJy. They used the
ILM with $\eta=1$ to interpret their data, and derived diameter and
albedo values of $D = 609$ km (562 -- 702 km), and $p_V = 0.05$ (0.04
-- 0.06).  Extrapolating our STM and ILM models to 1.2~mm we predict the
flux density there to be $0.6 \pm 0.16$~mJy, or 55\% of their observed
flux, although our $1\sigma$ upper bound of 0.76~mJy is not far from
their $1\sigma$ lower bound of 0.84~mJy. A similar discrepancy exists
between the \Spitzer\ results for (55565) 2002 AW$_{197}$ (Cruikshank
\etal\ 2004) and the sub-mm results reported by Margot \etal\  (2002),
although the disagreement in the case of \tc\ is somewhat worse. 

The Altenhoff \etal\ data for \tc\ were taken in queue mode in several
epochs during the winter months of 2001 -- 2003. Their errors include a
15\% absolute calibration uncertainty, and an 18\% uncertainty based
on the measured noise in the data. When data from all epochs are
stitched together, the signal integrates upward smoothly, in-spite
of being obtained in such a distributed way: there is nothing within
the data set to suggest that there is an issue (W.J. Altenhoff, {\it
priv. communication}). Nevertheless, it seems possible that combining
data from so many epochs might contribute to errors higher than nominal,
so perhaps the discrepancy with our data can be attributed to somewhat
optimistic error estimates for the 1.2~mm data.  If irreconcilable
differences between sub-mm measurements and Spitzer measurements turn
up for additional distant solar-system objects, a coordinated effort to
compare the two calibrations may be called for.

\subsection{Component diameters.}

Given the constraints of the observed magnitude difference of the pair,
$\Delta m$, from Margot \etal\  (2004; 2005b, and Table~1) and total brightness
consistent with our values for the effective diameter and albedo for the
system, some limits can be inferred on the sizes of the components of
\tc. Figure~4 shows the combinations that satisfy these constraints for
$D= 405$~km and $p_V = 0.079$ (these same basic curves are traced-out
for other values of $D$ and $p_V$, they simply extend to somewhat
larger or smaller values of $D$).  Assuming that the components have
equal albedos, the ratio of their diameters in terms of their observed
magnitude difference is $(d_1/d_2)^2 = 10^{\Delta m / 2.5}$, where $d_1$
is the radius of the primary. Given the effective diameter, $D$, of the
pair from our thermal results, the diameter of the primary is $d_1 =
D / (1 + (d_2/d_1)^2 )^{1/2}$, and $d_2$ is found from the previous
equation. Taking $D=405$~km, the sizes of equal-albedo components are 379
and 133 km. %(XXX as noted above, Margot \etal\ (2005a; 2005b) report similar
%visible colors for the components, so their albedos may also be similar XXX). 
An alternate case is that the 2 components have diameters
equal to $0.5 D$, and the apparent brightness difference is entirely
due to an albedo difference. In this case the above expressions can be
re-written to show that the albedos of the primary and companion are 0.28
and 0.039, respectively.  The most extreme sizes that are allowed are $d_1
= D = 405$~km (in which case $p_{V1}=0.069$), and $d_2 = 40$~km (in which
case  $p_{V2}=1$). An albedo of 1 is probably quite unrealistic
for a smallish KBO: a more reasonable upper limit might be $p_{V2} =
0.4$, in which case $d_2 = 66$~km.  The data collected by Margot \etal\ 
(2005b) were unsuitable for making the difficult measurement of the
mass ratio, but knowledge of that ratio would provide another useful
constraint on the sizes of the components, and thereby improve our
density determination (as discussed below).  In summary, for our best
effective diameter, $D = 405$~km, the component diameters seem likely to
lie in the ranges $202.5 \le d_1 \le 405$ and $202.5 \ge d_2 \ge 66$.

\section{Density}

Given the mass determined by Margot \etal\  (2004; 2005a; 2005b) and
our determination of the effective diameter of the \tc\ pair, the system
density in terms of the sizes of the primary and companion is
$$
\rho = {{6M}\over{\pi D^3}}\, {{(1+\Delta_2)^{3/2}}\over{1+\Delta_2^{3/2}}}
$$
where $M$ is the system mass, $\Delta_2 = 10^{-\Delta m/2.5} =
d_2^2/d_1^2$, and other quantities are as defined earlier.  Due to the
$D^3$ term, the error in our determination of the size of \tc\ strongly
dominates the error in the mass in determining the uncertainty in the
density. Table~4 summarizes the resulting densities for our adopted range
of $D$ and $p_V$, and for the range of component diameters discussed
above.  Considering the densities that result from the entire suite of
model parameters, we adopt $\rho = 0.5$~\gcc\ as our best estimate of the
density, with the true value likely to fall in the range 0.3 -- 0.8~\gcc.

Very low densities have been determined for many primitive solar system
objects, in particular small bodies.  For example, Asphaug and Benz
(1996) estimated the density of comet Shoemaker-Levy 9 to have been 0.6~\gcc\
based on observations of its tidal break-up during a close encounter with
Jupiter. Three main-belt asteroids with diameters $>$ 100~km have densities
$<$ 1.5~\gcc\ (15 Eunomia, 45 Eugenia, and perhaps 90 Antiope).  Of these, Eunomia
is particularly interesting because its density ($0.96 \pm 0.3$~\gcc) is
far below the average density for S-class (stony) asteroids (2.7~\gcc),
and because it is the largest asteroid (diameter 255~km) with such a
low density (see Britt \etal\ (2002) and Hilton (2002) and references
therein). Anderson \etal\ (2005) find that the density of Jupiter's moon
Amalthea (diameter $\simeq 83$~km) is $0.86\pm0.2$~\gcc. Margot and Brown 
(2001) discovered that 87 Sylvia was a binary, and using the IRAS diameter
derived a density of $\simeq1.6$~\gcc.  Marchis \etal\ (2005) subsequently
discovered that Sylvia is actually a triple asteroid (the first known,
with two satellites orbiting the primary), and estimated the density of
the $\simeq 150$~km diameter primary be $1.2\pm 0.1$~\gcc.  Trilling and
Bernstein (2005) applied rotational stability arguments to the 33 KBOs and
Centaurs with published lightcurves and show that none require densities
larger than 0.5 - 1.5 g/cc (for various models) to remain gravitationally
bound given their observed rotation rates. While this does not mean
their densities are in that range (higher densities are not ruled out),
it is interesting to note that the low density we find for \tc\ is in
accord with the lower-end of their limits. Because the rotation period
of \tc\ is unknown, their analysis can not be directly applied to it
(although such an analysis could be very revealing).

It is relatively easy to imagine strength-dominated objects less than
around 200 kilometers across having very low densities (and presumably
high porosities). Most of the low-density objects just mentioned are known
to have irregular shapes (see references given above), indicating material
strength does indeed dominate gravity in their interiors. Such densities
could be achieved via catastrophic disruption and re-accumulation
({\it e.g.} Richardson \etal\ 2002).  Such configurations are far less
intuitively appealing in the case of an object such as \tc, likely to
be more than 300 kilometers in diameter, in which gravitational forces
are likely to dominate material strength for some significant portion
of the interior.

\subsection{Interior Structure.}

Given an assumption for the density, $\rho_0$, of the material making up
the solid portions of \tc, the fractional void space, or porosity, $f$,
can be calculated from the total mass and the component diameters as $f =
1-\rho/\rho_0 = 1-(6 M)/(\pi (d_1^3+d_2^3)\rho_0)$.  Plausible values for
$\rho_0$ range from around 0.9~\gcc\  (that of water ice, almost certainly the
dominant constituent (\eg\ Anders \& Grevesse 1989), and other molecular
ices likely to be present) to that of Pluto and Triton (2~\gcc, McKinnon
\etal\ 1997), which are composed of roughly equal mass fractions of water
ice and ``rocky'' material. Pluto is a particularly likely compositional
surrogate for \tc, because both bodies occupy the same orbital resonance
with Neptune and so probably formed at a similar heliocentric distance
(although \tc\ has a lower orbital inclination, and so may have formed
exterior to Pluto).  While Pluto is differentiated, significantly
compressed by its own gravity, and probably lost some icy material in
a Charon-forming impact (Canup 2005) it is very difficult to imagine how \tc\
could have formed with a significantly different rock fraction. That being
said, our density for \tc\ would be easier to accept if $\rho_0$ were
near or even below 1~\gcc, suggesting a dearth of silicate materials, and
a corresponding enrichment of water ice and other low-density molecular
ices. Candidate ices other than water that might be somewhat abundant
are $CO$ ($\rho = 1.0$~\gcc), $N_2$ ($\rho = 0.95$~\gcc), and $CH_4$
($\rho = 0.5$~\gcc) (Scott, 1976; Jiang \etal\  1975; Manzhelii and
Tolkachev, 1964). However, appealing to such an extreme composition is
ad-hoc at best, and it seems highly unlikely that the low density can
be explained without significant internal porosity or some other effect.

Figure~5 shows the porosity we derive for \tc\ for the plausible range of
material densities, and for our adopted range of effective diameter for
the system.  Porosity values for selected values of $\rho_0$ are given
in Table~4. If we assume the material density is $1.2 < \rho_0
< 1.8$, then for an effective diameter of 405 km the range of porosity
required is about 62 -- 74\%. If we take 1.5~\gcc\ as a likely average
material density, then our adopted size limits constrain the porosity
to be between 48 -- 80\%. Porosities in the range 45 -- 80\% cover most
of the allowed values in Figure~5.  If we take $D = 609$~km from
from Altenhoff \etal\ (2004) and set $\rho_0 = 0.8$~\gcc, the porosity
is 83\%, and it gets progressively higher for larger material densities.
We conclude that our results, if interpreted in terms of void spaces 
internal to \tc, require the porosity to be about 65\%, with a range of
about 45 -- 80\%.  It is astonishing to think of such a large body having
more than half of its interior volume taken up by voids. Such an object
strains the bound of what might reasonably be called a ``rubble-pile''
(Weissman 1986), being volumetrically more akin to a ``void-pile'',
with some solid matter thrown in.

Our lower porosity limit, 45\%, is still well in excess of the 26\%
expected for closely-packed equal-sized spheres, and is also in excess
of that expected for randomly-packed equal-sized spheres (36\%, \eg\
Torquato \etal\ 2000). If the component pieces are unequal, porosity
decreases because the smaller ones will infill the gaps between the
larger ones. An important effect that increases the porosity of a
collection of particles is friction, which could be due to roughness
and/or angularity of those particles.  Thus, a model which might be
consistent with the high porosity implied by our density measurement is
that the rubble in the interior of \tc\ be both very irregular and of
comparable size.  Another could be that the rubble is itself porous. Such
a configuration might result from relatively gentle assembly of grains to
make macro particles, which then assemble to form porous ``boulders'',
and so-on until a large object with significant porosity at all scales
results. Similarly, sublimation of volatile components from within a
water ice matrix could result in porosity at multiple scales. However,
such paradigms ignore the fact that considerable momentum and energy
must be dissipated during accretion, because accreted material will
be deposited {\it via} high-velocity impacts. Such impacts will lead
to local compression, melting, and vaporization, and in some cases
large-scale fracturing ({\it e.g.} Asphaug \etal\ 1996; Richardson \etal\
2002). Indeed, impacts would generally tend to compress such a porous
structure as that discussed above, casting doubt on whether it could
arise during accretion or be maintained during subsequent billions of
years of cratering.

\tc\ is also large enough that the expected internal stresses should
lead to crushing and densification in the interior.  The stress at the
center of a homogenous sphere is $P_0 = \pi G \rho^2 D^2 / 6$. Taking
the highest-pressure scenario for \tc\ we set $\rho = 1$ (somewhat
above the top of our range for density) and $D = 405$~km and find
$P_0 = 6\times10^7$~dyn/cm$^2$.  The yield strength of water ice is
$10^7$~dyn/cm$^2$ (Beeman, 1988) at low confining pressures. In the case
of a rubble-pile object, the confining pressure is essentially zero ({\it
i.e.} the entire weight of overburden is comparable to the deviatoric
stress at contact points between the fragments), and the low-pressure
strength is the relevant one. So if the density of \tc\ were around
1~\gcc\ and the primary were as large as our best-fit effective diameter
for the system, voids in the deep interior would indeed be closed via
crushing of icy material. However, $P_0$ depends on the square of both
the density and the radius, so modest reductions in either or both lead
to significant reduction in interior stress. For a density of 0.5~\gcc\
and a diameter of 300~km, $P_0 = 8\times10^6$~dyn/cm$^2$. It appears
that \tc\ may be on the verge of being large and dense enough that it
would be compressed, at least near the center, by its own weight. If so,
and if the density of larger KBOs can be measured, we predict that the
larger objects would indeed have to be denser.

The finding that significant strength-supported void space could exist in
the interior of \tc, is still somewhat puzzling.  The lack of a lightcurve
greater than 5 -- 10\% (Peixinho \etal\  2002, Ortiz \etal\  2003;
Margot \etal\  2005a) is itself evidence that gravity dominates strength
in the interior, resulting in the relaxation of meso-scale topography
and a near-spherical shape. To explore this apparent inconsistency,
we examined the possibility that \tc\ has a low density (porous) outer
mantle, and a denser, non-porous core.  Such a configuration might arise
in an impact that produced a largest fragment that was a significant
fraction of the size of the original body, as would be expected from
a collision with a specific energy at or somewhat above the shattering
specific energy, $Q^*$ ({\it e.g.} Holsapple \etal., 2002).  Figure~6
shows results for the average density and porosity of \tc\ as a function
of the size of a hypothetical core. The figure assumes that the core is
composed of a rocky component, with densities taken from the McKinnon
\etal\ (1997 and references therein) models for the composition of
Pluto, and that the density of the mantle is 0.5~\gcc\ (consistent with
a composition dominated by water-ice with about 50\% porosity). The
upper-limit on the density of the binary, 0.8~\gcc, places only a weak
constraint on the size of a rocky core within \tc: it could be as large
as 0.45--0.51 of the total radius (depending on the core composition)
and still satisfy our upper-bound on the density of \tc. If the density
is really 0.5~\gcc, there can be no core unless the mantle density is
still lower. Setting the mantle density to 0.3~\gcc ({\it i.e.} setting
its porosity to 70\%), we find that a rocky core could exist within \tc\,
but that its size would be limited to 0.40--0.45 of the total radius.
These results offer a compromise wherein \tc\ has a relatively dense
core overlain by a very porous, water-ice mantle. The sperical shape of
the core would presumably help to moderate the external form, causing
the overall shape to tend towards the sphericity suggested by the lack
of a measureable lightcurve.

In summary, while the nominal density and porosity we determined above
are not strictly inconsistent with considerations of cosmochemistry,
formation, evolution, and strength {\it vs.} gravity, in each of these
areas it seems that our nominal density range (0.3--0.8~\gcc) is extreme. 
The tolerance of our density for the presence of a high-density core,
even if that core is dominated by silicates with densities around 
3~\gcc, provides for a plausible scenario where \tc\ can have a 
low-density/high-porosity and also have a spherical shape. However,
such a low density for such a large object is still extreme.

\subsection{Multiple System?}

An alternative to the porous/rubble-pile models above is that one or
both components of \tc\ are very irregular in external form, essentially
incorporating significant void space within their apparent limbs. As
just discussed, this seems unlikely because of the lack of a measured
lightcurve. However, an extreme case of such external porosity would be
if one or both of the two known components are also binary (or multiple)
systems. Three multiple minor-planet systems are known: 87 Sylvia is
a triple (Marchis \etal\ 2005), Pluto is a quadruple (Weaver \etal\
2005; Buie \etal\ 2005) and 2003 EL$_{61}$ is a triple (Brown \etal\
2005). If \tc\ is a triple, the lack of a lightcurve is not particularly
telling, as such a sub-binary would only possesses a lightcurve when
one component eclipses the other (during mutual events), assuming the
intrinsic lightcurve of each component is negligible. For a multiple
system of $N$ equal-size components, the density is given by $\rho =
6 \sqrt{N} M / (\pi D^3)$: the density for the multi-component system is
enhanced by a factor of $\sqrt{N}$ relative to that for a single component
system. In the case of \tc\ we postulate that the primary may itself be
an unresolved binary with component diameters of $d_1$ and $d_3$; the
diameter of the resolved companion to the primary is $d_2$, as before.
If we restrict our consideration to the case where the three components
have equal albedos, the density can be written as
$$ \rho = {{6M}\over{\pi D^3}}\, {{[(1+\Delta_2)(2+\Delta_3)]^{3/2}} \over
                     {1+\Delta_3^{3/2}+\Delta_2^{3/2}(1+\Delta_3)^{3/2}}}
$$ 
where $\Delta_2 = d_2^2/(d_1^2 + d_3^2)$, $\Delta_3 = d_3^2/d_1^2$.
In addition, $\Delta_2 = 10^{-\Delta m / 2.5} = 0.139$ from the observed
magnitude difference between the primary and secondary.

Figure~7 shows the second term of the above expression as a function 
of the ratio $d_1/d_3$. From the figure it is
clear that to have a significant effect on the density we derive, the 
primary must be split into components of comparable size.  If the
\tc\ primary is actually 2 equal-sized components, this shows that the
density of a trinary \tc\ is enhanced by a factor of 1.60 relative to
the density for a single-body, and 1.38 relative to the density of a
binary sytem ($d_3 = 0$).  The resulting average density of the triple 
system is 0.7~\gcc, with a range of 0.4 -- 1.1~\gcc\ (compared to 0.3 --
0.8~\gcc\ for the binary system). The porosities derived earlier decrease
to about 0.5 with a range of 0.1 -- 0.7. Table~4 summarizes our results
for the density of this hypothetical trinary system.  These densities
and porosities might be achievable given the likely major constituents
of \tc, and realizable internal structures. Somewhat larger increases
to the density could result if the secondary has a low albedo and is
therefore comparable in size to the primary. In this case, and in the
unlikely event that the secondary is also double, the density could be
enhanced by a factor approaching 2 relative to the single-body density,
or by a factor approaching 1.7 relative to the equal-albedo binary case.

\section{Conclusions}

We detected thermal emission from the binary KBO \tc\ at 24 and 70~\mum\
using the \Spitzer\ space telescope. When interpreted using a range
of thermal models, we derive an effective diameter for the system in
the range $350 \leq D \leq 470$~km, with the best value being 405~km.
The corresponding range of the visible geometric albedo is $0.055
\leq p_v \leq 0.11$, with the best value being 0.079. When combined
with the system mass determined from HST data by Margot \etal\ (2004;
2005a; 2005b), our size determination indicates an average density in
the range 0.3 -- 0.8~\gcc.  For likely bulk compositions, dominated
by roughly equal mass fractions of rocky material and water ice, the
porosity required to explain densities this low is in the range 55 --
75\%. Such high porosities strain the bounds of what might reasonably
be expected for naturally occurring internal structures for such a large
object. However, we do find that \tc\ is just small enough that it might
not be compressed under its own weight, so porosities this high can
not be easily ruled out on those grounds. It is possible that the \tc\
primary could have a core with a density around 3~\gcc, so long as that
core is smaller than about 1/2 the radius. The presence of a dense core
may help reconcile the apparent inconsistency between the presence of
large amounts of strength-supported void space in the interior and the
nearly spherical shape evidenced by the lack of a measured lightcurve.

If \tc\ is actually a triple system, the densities and porosities
we derive must be modified accordingly. We derive a general expression
for the density of the system as a function of the component sizes, and
show that the density could actually be as much as 1.38 times greater than
for the binary system, if the primary is split into comparably-sized
components. Such a trinary system would be consistent with the lack of
an observed lightcurve so long as none of the components eclipse one
another (if the orbit of the primary-binary is inclined 49\degr\ to our
line of sight, the centers of equal-sized components could be as close
as three radii without eclipses occurring). The density of such a system
would be in the range 0.4 -- 1.1~\gcc.  At the upper end, these densities
are comparable to that of, \eg, Saturn's moons Tethys and Iapetus (\eg\
Burns 1986).

These results suggest that in addition to being a binary KBO, \tc\
has unexpected internal properties. The large model uncertainties
in the density and porosity could be reduced with additional data
that provided constraints on the individual sizes of the primary and
secondary, or that constrained their individual masses.  In addition,
very high-resolution imaging or lightcurve monitoring could help resolve
the question of whether the \tc\ primary is itself actually multiple.
If \tc\ were found to have a short rotation period, that could place a
lower bound on the density for which it would be stable against distortion
or breakup due to rotation.

\acknowledgments
This work is based [in part] on observations made with the Spitzer
Space Telescope, which is operated by the Jet Propulsion Laboratory,
California Institute of Technology under a contract with NASA. Support
for this work was provided by NASA through an award issued by JPL/Caltech.

%%%%%%%%%%%%%%%%%%%%% BIBLIOGRAPHY %%%%%%%%%%%%%%%%%%%%%%%%%%%%%%%%%%%%%%%%%%%%%%%%%%

%%%%%%%%%%%%%%%%%%%%% FIGURES %%%%%%%%%%%%%%%%%%%%%%%%%%%%%%%%%%%%%%%%%%%%%%%%%%%%%%%

%% Figure 1 %%%%%%%%%%%%%
\clearpage
\begin{figure}
\epsscale{1.0}
%\plotone{99TC36_2band.eps}
\plotone{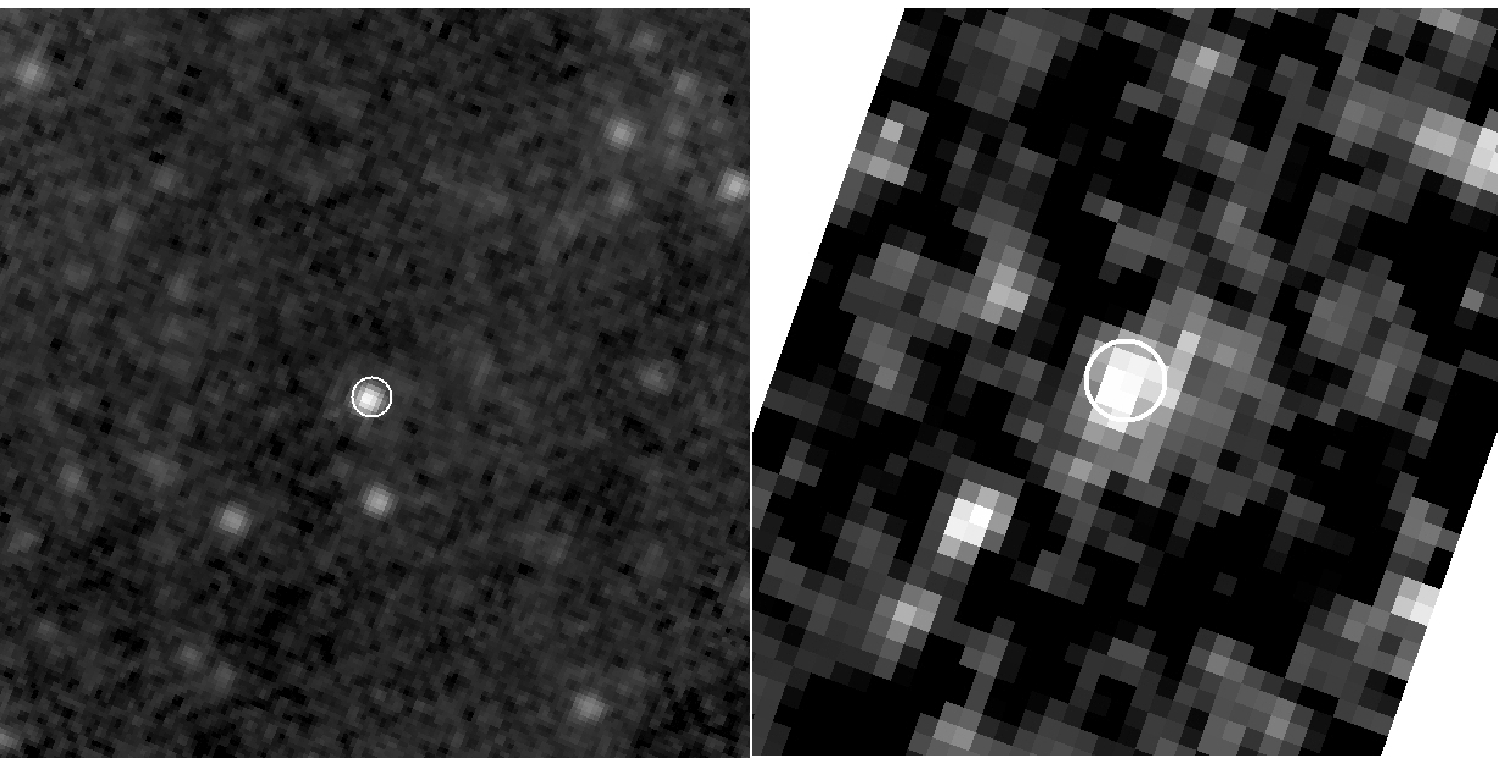}
\caption{
Images of \tc\ at 24 \mum\ (left) and 70 \mum\ (right). Each image is $190\arcsec$
square, and the orientation is North up, East left. The circles are centered at the
ephemeris position of the target. It is just possible to make out the first Airy
maximum in the 24~\mum\ image. There is no significant background structure due to cirrus
at either wavelength.
\label{fig1}
}
\end{figure}

%% Figure 2 %%%%%%%%%%%%%
\clearpage
\begin{figure}
\epsscale{0.7}
%\plotone{99TC36_SEDs.eps}
\plotone{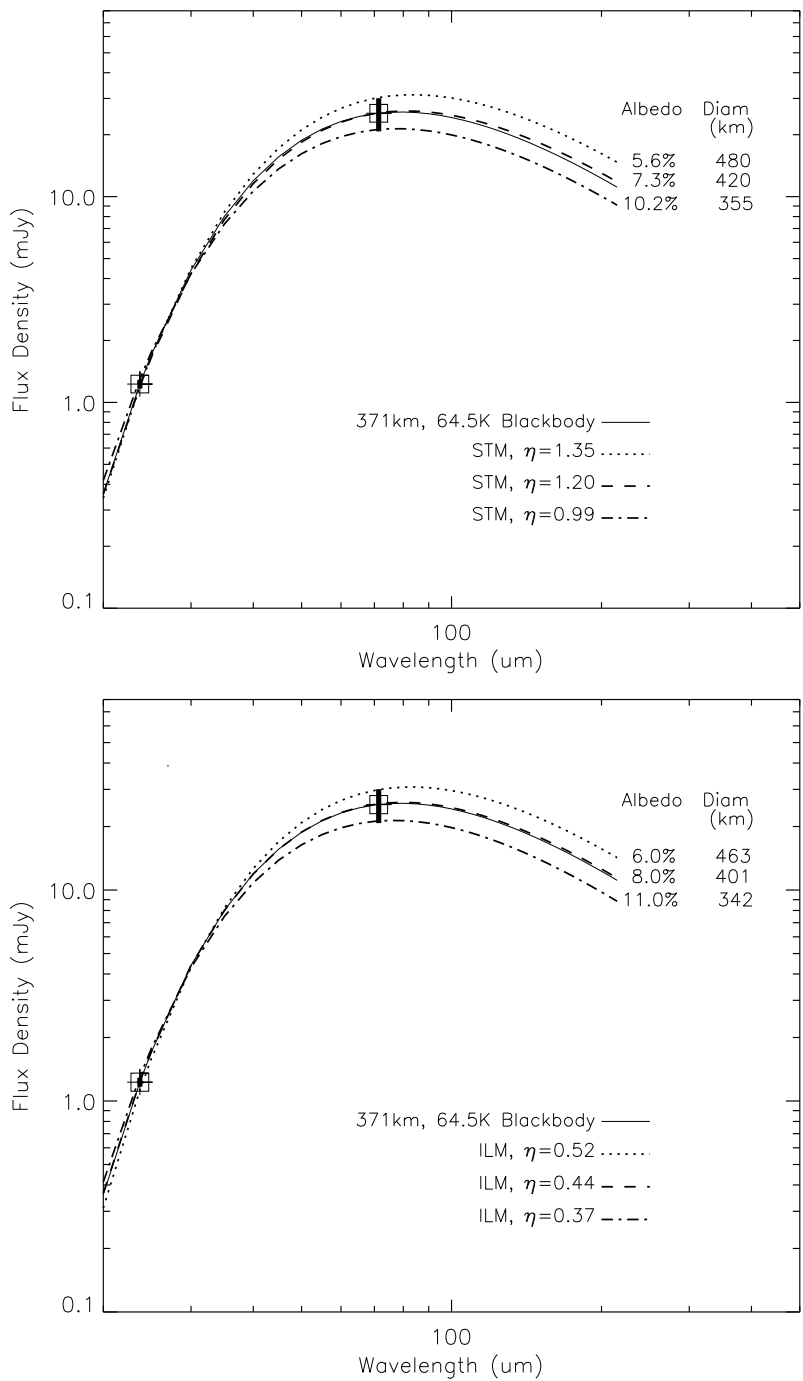}
\caption{
Thermal models fitted to our 24 and 70 \mum\ photometry. Results from
the Standard Thermal Model (STM) are given in panel a (top), and those
from the Isothermal Latitude model (ILM) in panel b (bottom). Diameters,
which are the effective total diameter for both components of the binary
system, and effective geometric albedos corresponding to each model are
given at upper right. The beaming parameter, $\eta$, for each model is
given in the legend. The temperature of a zero-albedo surface at the
distance of \tc\ would be 70.6~K.
\label{fig2}
}
\end{figure}

%% Figure 3 %%%%%%%%%%%%%
\clearpage
\begin{figure}
\epsscale{0.5}
%\plotone{TPhysPlot.eps}
\plotone{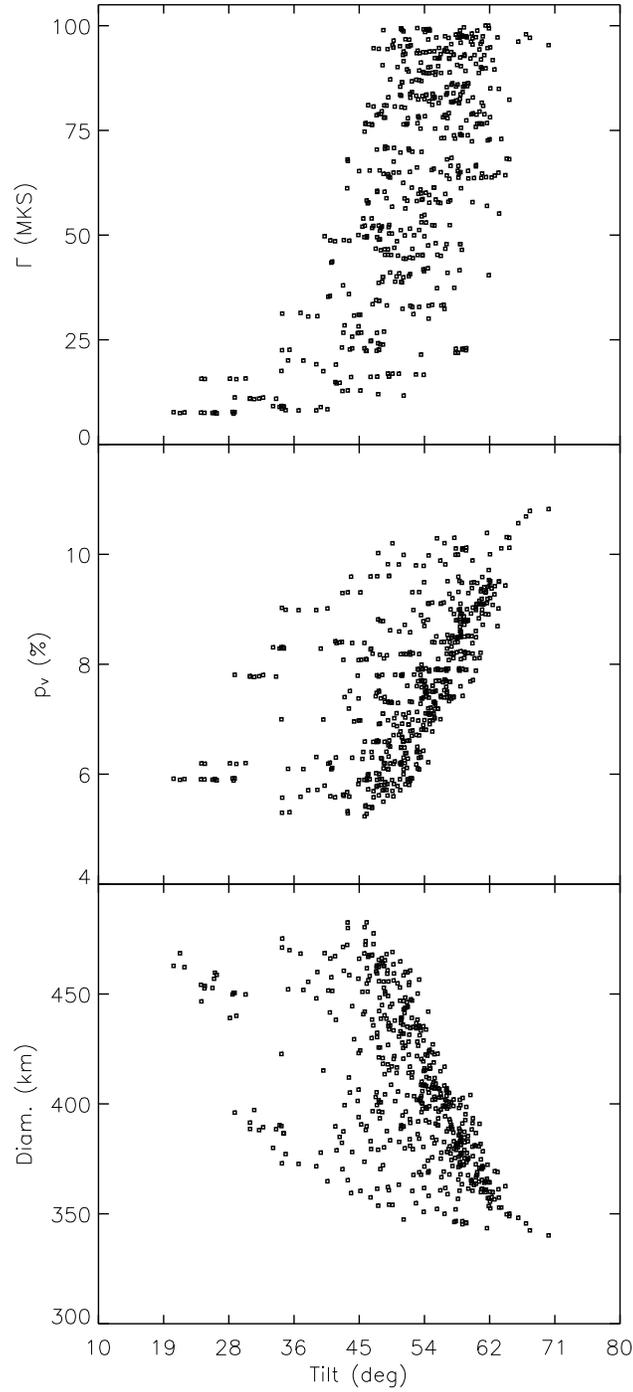}
\caption{
Model parameters for rough-surface thermophysical models consistent
with our thermal photometry and the absolute V magnitude of \tc. A
point is plotted for each of the 539 models that fit the data (of
30000 models run).
\label{fig3}
}
\end{figure}

%% Figure 4 %%%%%%%%%%%%%
\clearpage
\begin{figure}
\epsscale{1.0}
%\plotone{p_vs_d.eps}
\plotone{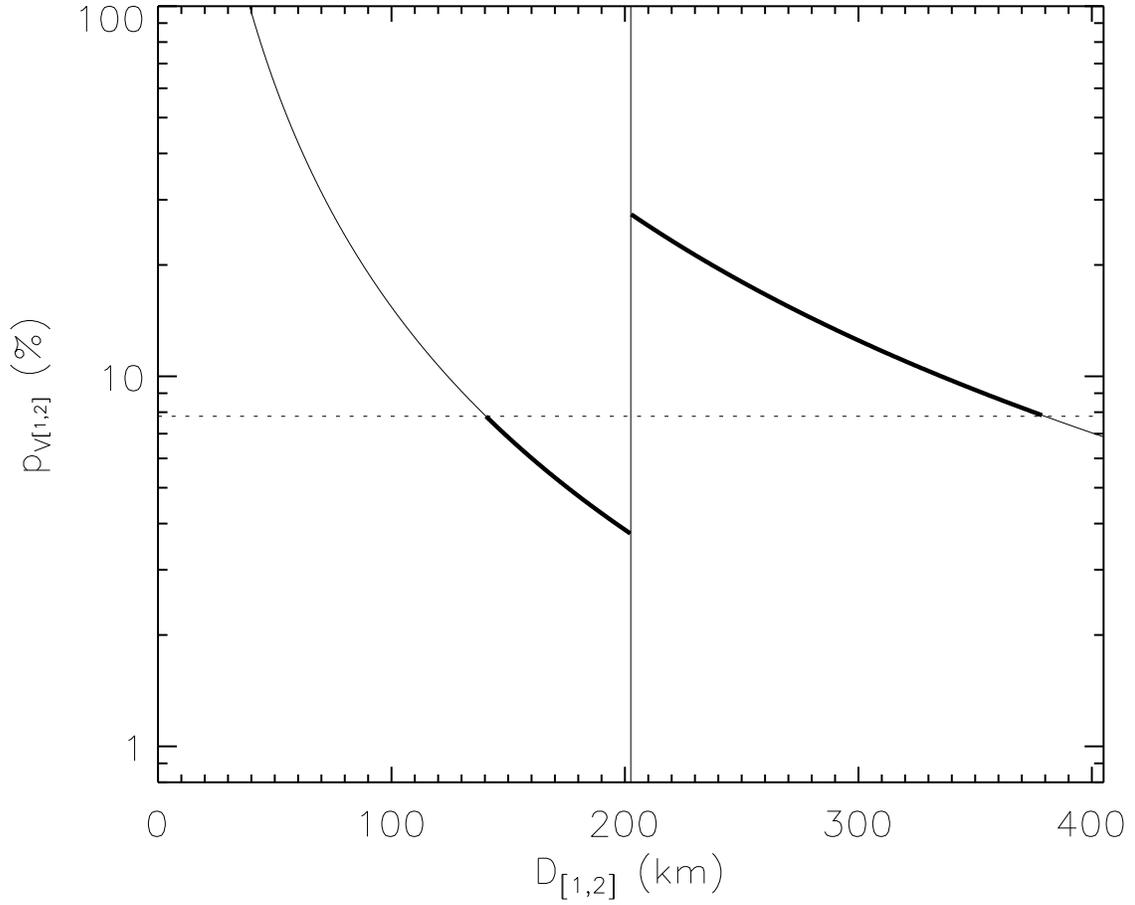}
\caption{
The locus of combinations of albedo and diameter for the \tc\ primary (to
the right of the vertical line), and companion. The solution shown applies
for an effective system diameter and albedo of 405 km and 7.9\%. The
heavy lines show the range of solutions where the primary has a higher
albedo than the secondary; the thin lines those ranges where the primary
has a lower albedo than the secondary. 
\label{fig4}
}
\end{figure}

%% Figure 5 %%%%%%%%%%%%%
\clearpage
\begin{figure}
\epsscale{1.0}
%\plotone{densplot2new.eps}
\plotone{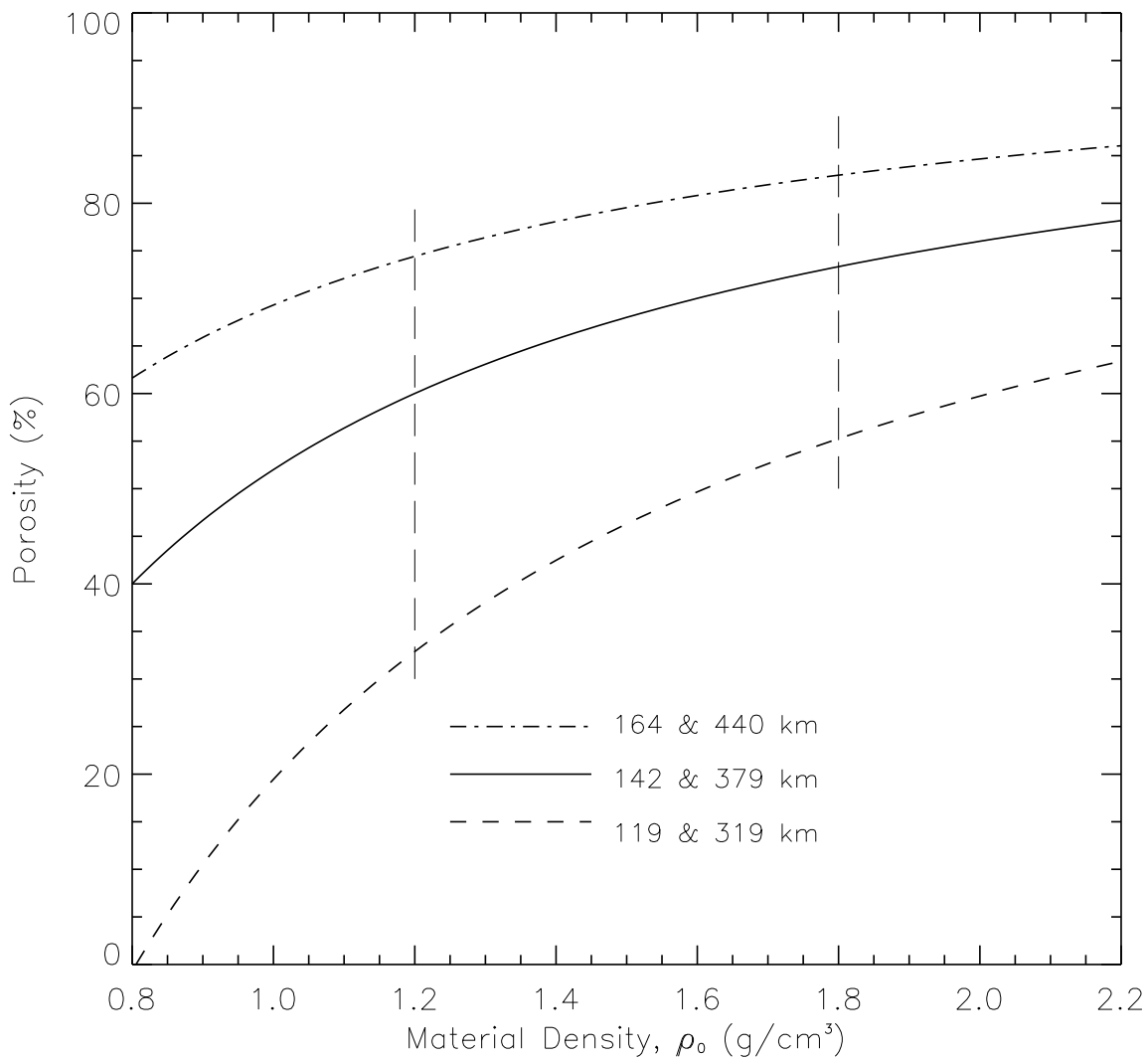}
\caption{
The fraction of void space within \tc\ resulting from our determination of the
effective diameter and the mass determination of Margot \etal\  (2005b). The 3 
lines give the dependence for our adopted value (405~km) and limits (350 -- 470~km)
for the effective diameter. The legend gives the corresponding sizes of the
components if they have equal albedos. The vertical lines indicate a reasonable
range of material density in the outer Solar System (see text).
\label{fig5}
}
\end{figure}

%% Figure 6 %%%%%%%%%%%%%
\clearpage
\begin{figure}
\epsscale{1.0}
%\plotone{dens_2layer.eps}
\plotone{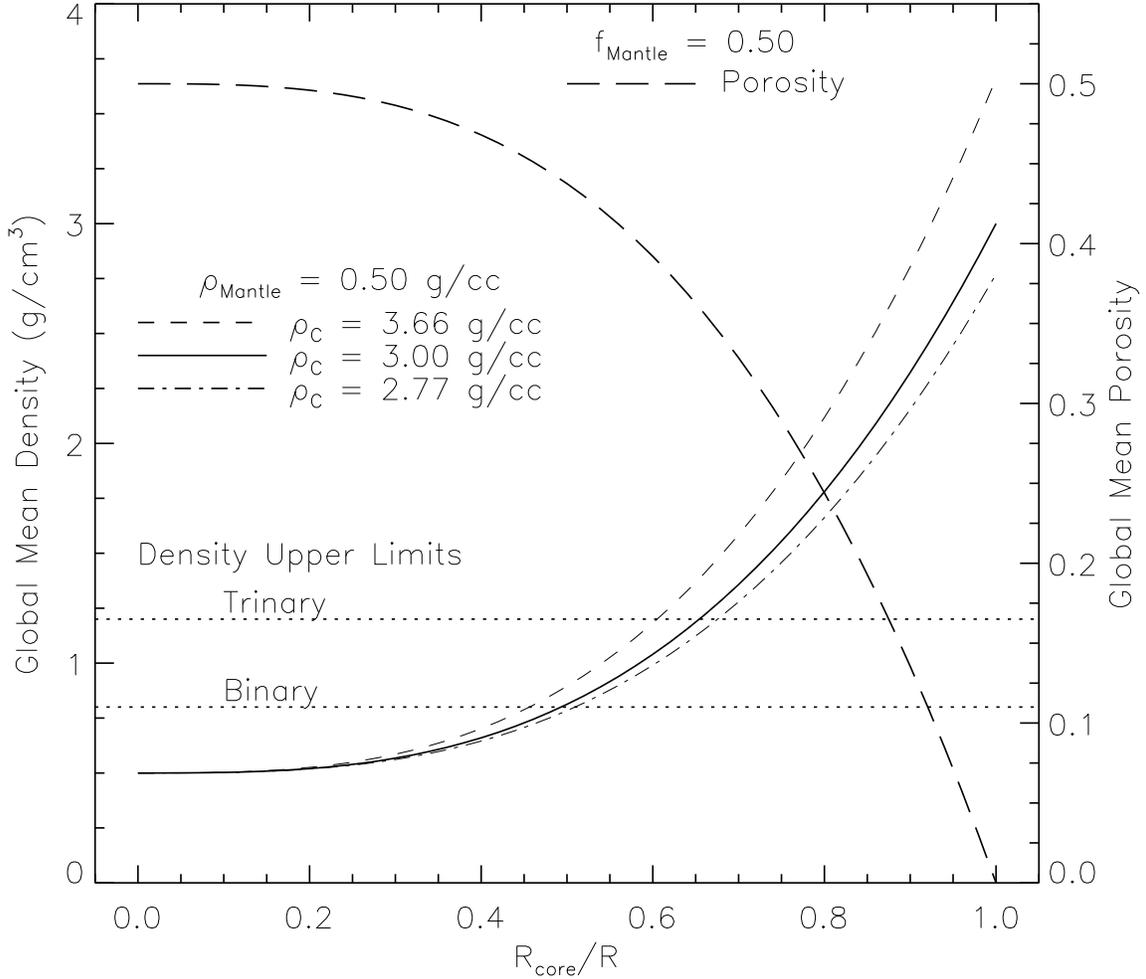}
\caption{
The density and porosity of \tc\ as a function of the size of a hypothetical
core. The density of the mantle (the layer surrounding the core) is taken to be 
0.5~\gcc, consistent with a composition dominated by water ice with 50\% porosity.
The density of the material in the core (assumed to be dominated by
silicates) is reflected by the legend labels $\rho_C$, with the resulting global
mean density indicated by the corresponding linestyles.  The average porosity
of the entire 2-layer structure is given by the long-dashed line. Our upper limits
on the density of binary and trinary versions of \tc\ are shown as horizontal
dotted lines. In spite of the high porosities required by our density, large
non-porous cores with densities appropriate for rocky material are consistent
with our data.
\label{fig6}
}
\end{figure}

%% Figure 7 %%%%%%%%%%%%%
\clearpage
\begin{figure}
\epsscale{1.0}
%\plotone{TriDens.eps}
\plotone{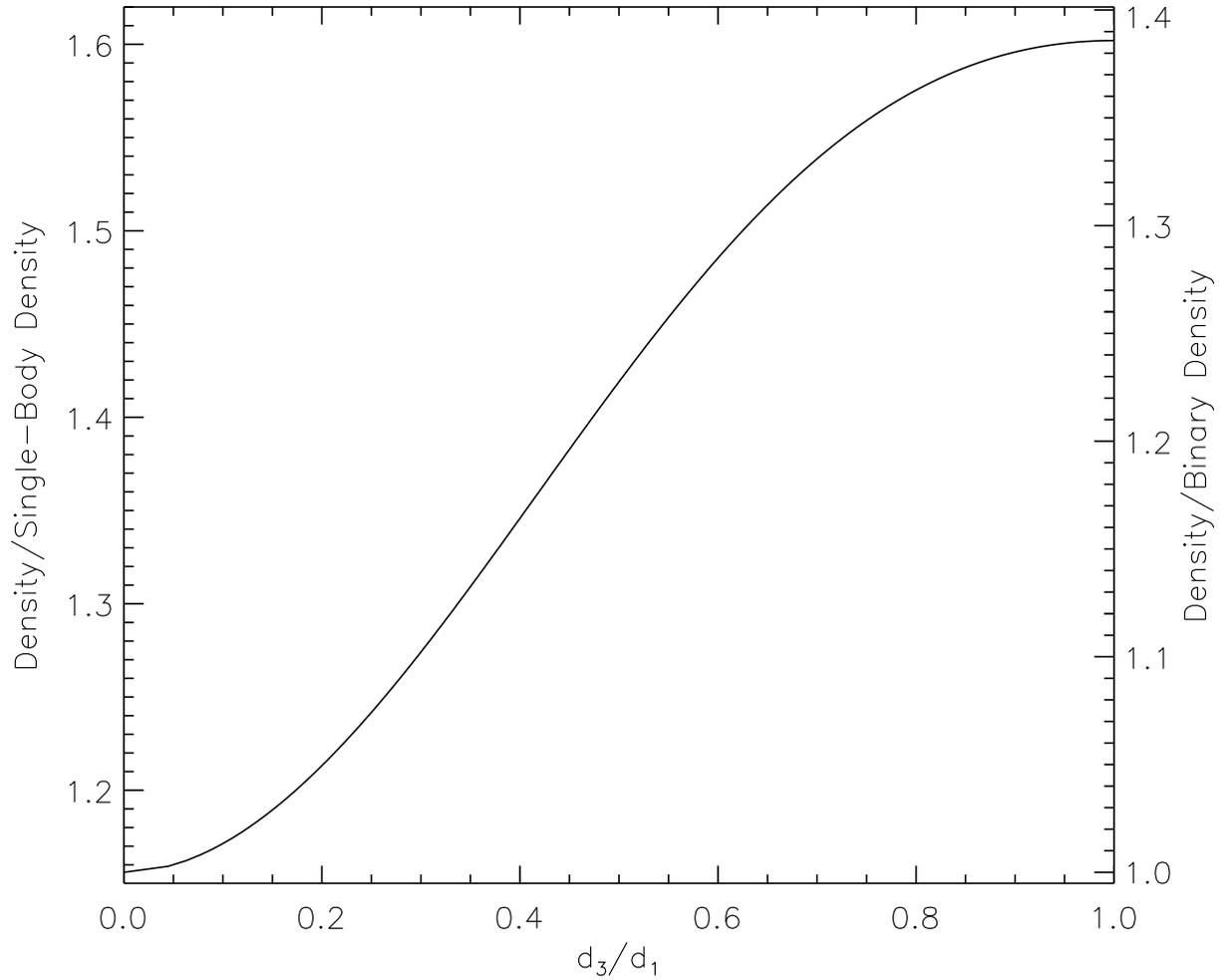}
\caption{
Density enhancement relative to the single-component system density
($6M/(\pi D^3)$) for a trinary system. The relationship shown assumes
that all 3 components have the same albedo. For $d_3/d_1 = 0$ the system
is a binary, and for $d_3/d_1 = 1$ the ``primary'' is itself a binary
with 2 equal-sized components.
\label{fig7}
}
\end{figure}

%%%%%%%%%%%%%%%%%%%%% TABLES %%%%%%%%%%%%%%%%%%%%%%%%%%%%%%%%%%%%%%%%%%%%%%%%%%%%%%%%

%% Table 1 %%%%%%%%%%
\begin{deluxetable}{lcc}
\tabletypesize{\footnotesize}
\tablecaption{The 1999 TC$_{36}$ Binary System \label{tbl1}}
\tablewidth{0pt}
\tablehead{
 \colhead{Parameter} & \colhead{Symbol} & \colhead{Value}
}
\startdata
 {\it ~~Heliocentric Orbit} \\
 semi-major axis & $A$\tablenotemark{a}        & 39.23 AU \\
 eccentricity    & $e$\tablenotemark{a}        & 0.22 \\
 inclination     & $i$\tablenotemark{a}        & $8.4\degr$ \\
 {\it ~~Binary System} \\
 semi-major axis & $a$\tablenotemark{b}        & 7720 km \\
 system mass     & $M$\tablenotemark{b}        & $1.44\times10^{19}$ kg \\
 orbit period    & $P$\tablenotemark{b}        & 50.36 d \\
 contrast        & $\Delta m$\tablenotemark{b}\tablenotemark{c}
                                               & $2.14\pm 0.02$ \\
 size ratio      & $r_1/r_2$\tablenotemark{d}  & 2.68 \\
\enddata

\tablenotetext{a}{Heliocentric orbit values from the Minor Planet Center.}
\tablenotetext{b}{Values from Margot \etal\ 2004; 2005a; 2005b. The uncertainty
in the mass is about 15\%.}
\tablenotetext{c}{Visual magnitude difference between primary and secondary.}
\tablenotetext{d}{Radius ratio of the components assuming equal albedos.}
\end{deluxetable}

%% Table 2 %%%%%%%%%%
\begin{deluxetable}{cccccccc}
\tabletypesize{\footnotesize}
\tablecaption{Observational Circumstances and Flux Densities \label{tbl2}}
\tablewidth{0pt}
\tablehead{
\colhead{Wavelength} & \colhead{}     & \colhead{Exposure} & \colhead{$R_{Sun}$}
          & \colhead{$\Delta_{Spitzer}$} & \colhead{Flux (error)}\\
\colhead{($\mu$m)}   & \colhead{Date\tablenotemark{a}} & \colhead{(sec)}    & \colhead{(AU)}                                                                                              
          & \colhead{(AU)}               & \colhead{(mJy)}
}
\startdata
24  & 2004 Jul 12 10:41 & 1400 & 31.10 & 30.94 &  1.23 (0.06)\tablenotemark{b} \\
70  & 2004 Jul 12 11:09 &  800 &  ...  &  ...  & 25.4 (4.7)\tablenotemark{b} \\
\enddata
                                                                                             
\tablenotetext{a}{The J2000 pointing for the observations was 00:45:29.8, -04 16 59.}
\tablenotetext{b}{Errors include those from the absolute calibration. The SNR of the
detections was $\simeq 50$ at 24~\mum, and $\simeq 10$ at 70~\mum.}
\end{deluxetable}

%% Table 3 %%%%%%%%%%
\begin{deluxetable}{lccccccc}
\tablecaption{Thermal Model Results\label{tbl3}}
\tablewidth{0pt}
\tablehead{
\colhead{Model} &
\colhead{$D$ (km)\tablenotemark{a}} &
\colhead{$p_V$ (\%)\tablenotemark{a}} &
\colhead{$\eta$\tablenotemark{a}} &
\colhead{$\Gamma$\tablenotemark{a}\tablenotemark{c}}
}
\startdata
STM            & 420 (355 -- 480) & 7.3 (10 -- 5.6)  & 1.2 (1.0 -- 1.4)    & 0 \\
ILM            & 401 (342 -- 463) & 8.0 (11 -- 6.0)  & 0.44 (0.37 -- 0.52) & $\infty$ \\
ILM (tilted)\tablenotemark{b}
               & 420 (364 -- 488) & 7.3 (9.7 -- 5.4) & 0.80 (0.7 -- 0.95)  & $\infty$ \\
T.phys. (10hr)\tablenotemark{b}
               & 434 (362 -- 504) & 6.9 (9.9 -- 5.1) & ... & 3.5 (0.3 -- $\infty$) \\
T.phys. (40hr)\tablenotemark{b}
               & 434 (362 -- 504) & 6.9 (9.9 -- 5.1) & ... & 7.9 (0.7 -- $\infty$) \\
Adopted        & 405 (350 -- 470)  & 7.9 (11 -- 5.8) & ... & ... \\
\enddata
\tablenotetext{a}{Model results given as best value and (range).}
\tablenotetext{b}{The tilted ILM and thermophysical models assumed a sub-solar
and sub-\Spitzer\ latitude of 48.6\degr.}
\tablenotetext{c}{The units of $\Gamma$ are J\,m$^{-2}$\,K$^{-1}$\,s$^{-1/2}$.}
\end{deluxetable}

%% Table 4 %%%%%%%%%%
\begin{deluxetable}{lccccccc}
\tablecaption{Density Results\label{tbl4}}
\tablewidth{0pt}
\tablehead{
\colhead{} &
\colhead{} &
\colhead{} &
\colhead{Porosity\tablenotemark{a}\tablenotemark{b}} &
\colhead{} &
   \\
\colhead{Model} &
\colhead{Density (\gcc)\tablenotemark{a}\tablenotemark{b}} &
\colhead{($\rho_0 = 1.2$~gcc)} &
\colhead{($\rho_0 = 1.5$~gcc)} &
\colhead{($\rho_0 = 1.8$~gcc)} 
}
\startdata
\,\,{\it Binary} \\
Equal $p_v$      & 0.48 (0.81 -- 0.31) & 0.59 (0.39--0.72) & 0.67 (0.50--0.77) & 0.71 (0.58--0.80) \\
Equal Size (D/2) & 0.59 (0.99 -- 0.37) & 0.51 (0.26--0.68) & 0.60 (0.40--0.73) & 0.66 (0.49--0.77) \\
Max Difference   & 0.42 (0.73 -- 0.27) & 0.64 (0.45--0.75) & 0.70 (0.55--0.80) & 0.74 (0.62--0.83) \\
Adopted          & 0.5 (0.9 -- 0.3)    & 0.53 (0.32--0.72) & 0.65 (0.45--0.77) & 0.69 (0.54--0.80) \\[3pt]
{\,\,{\it Trinary}} \\
Equal Size\tablenotemark{c}\tablenotemark{d}
                 & 0.69 (1.10 -- 0.41) & 0.43 (0.08--0.66) & 0.54 (0.26--0.72) & 0.62 (0.39--0.77) \\
\enddata
\tablenotetext{a}{Results given as best value and (range).}
\tablenotetext{b}{The values correspond to our adopted values for $D$, with further 
constraints as noted in the Model column.}
\tablenotetext{c}{The trinary model assumes equal albedos for all 3 components. The
primary is assumed to be 2 equal-sized components ($d_3=d_1$).}
\tablenotetext{d}{Trinary model values are based on the adopted binary densities and the
density enhancement for $d_3=d_1$.}
\end{deluxetable}

\end{document}